\documentclass{aa}
\usepackage{psfig}

\usepackage[T1]{fontenc}
\usepackage{times}

\def\EXparamUno{}
\def\EXparamDue{}
\def\EXparamTre{}

\begin{document}
\thesaurus{     20(
		04.01.1;        
                04.03.1;        
                08.08.1;        
                08.16.3;        
                10.07.2)        
          }
\title {        Photometric catalog of nearby globular clusters (II).
\thanks {       Based on observations made with the 1m Jacobus Kapteyn
                Telescope operated on the island of La Palma by the
                ING in the Spanish Observatorio del Roque de Los
                Muchachos of the Instituto de Astrof\`{\i}sica de
                Canarias.}  }
\subtitle      {A large homogeneous $(V,I)$ color-magnitude diagram data-base.}
\author {       A. Rosenberg \inst{1}, 
                A. Aparicio  \inst{2},
                I. Saviane   \inst{3} \and 
                G. Piotto    \inst{3}}
\offprints {    Alfred Rosenberg: alf@ll.iac.es   }
\institute{     
                Telescopio Nazionale Galileo, 
                vicolo dell'Osservatorio 5, I--35122 Padova, Italy
\and
                Instituto de Astrofisica de Canarias, 
                Via Lactea, E-38200 La Laguna, Tenerife, Spain
\and
                Dipartimento di Astronomia, Univ. di Padova, 
                vicolo dell'Osservatorio 5, I--35122 Padova, Italy
        }
\date{}
\titlerunning {{\it VI} photometric catalog of nearby GGC's (II)}
\maketitle


\begin{abstract}
In this paper we present the second and final part of a large and
photometrically homogeneous CCD color-magnitude diagram (CMD) data
base, comprising 52 nearby Galactic globular clusters (GGC) imaged in
the $V$ and $I$ bands.

The catalog has been collected using only two telescopes (one for each
hemisphere). The observed clusters represent $75\%$ of the known
Galactic globulars with $(m-M)_{\rm V}\leq 16.15$~mag, cover most of
the globular cluster metallicity range ($-2.2 \leq {\rm [Fe/H]}\leq
-0.4$), and span Galactocentric distances from $\sim1.2$ to $\sim18.5$
kpc.

In particular, here we present the CMDs for the 13 GGCs observed in
the Northern hemisphere. The remaining 39 Southern hemisphere clusters
of the catalog have been presented in a companion paper (Rosenberg et
al. \cite{rosenberg00}). We present the first CCD color magnitude
diagram for NGC~6779 (M56).

All the CMDs extend from at least $\sim2$ magnitudes below the
turn-off (i.e. $V_{\rm lim}\geq22$) to the tip of the red giant
branch.  The calibration has been done using a large number of
standard stars, and the absolute calibration is reliable to a
$\sim0.02$~mag level in both filters.

This catalog, because of its homogeneity, is expected to represent a
useful data base for the measurement of the main absolute and relative
parameters characterizing the CMD of GGCs.

\keywords{Astronomical data base: miscellaneous - Catalogs - Stars:
Hertzsprung-Russell (HR) - Stars: population II - Globular clusters:
general}
\end{abstract}


\section{Introduction} 
\label{intro}

\begin{figure}
\psfig{figure=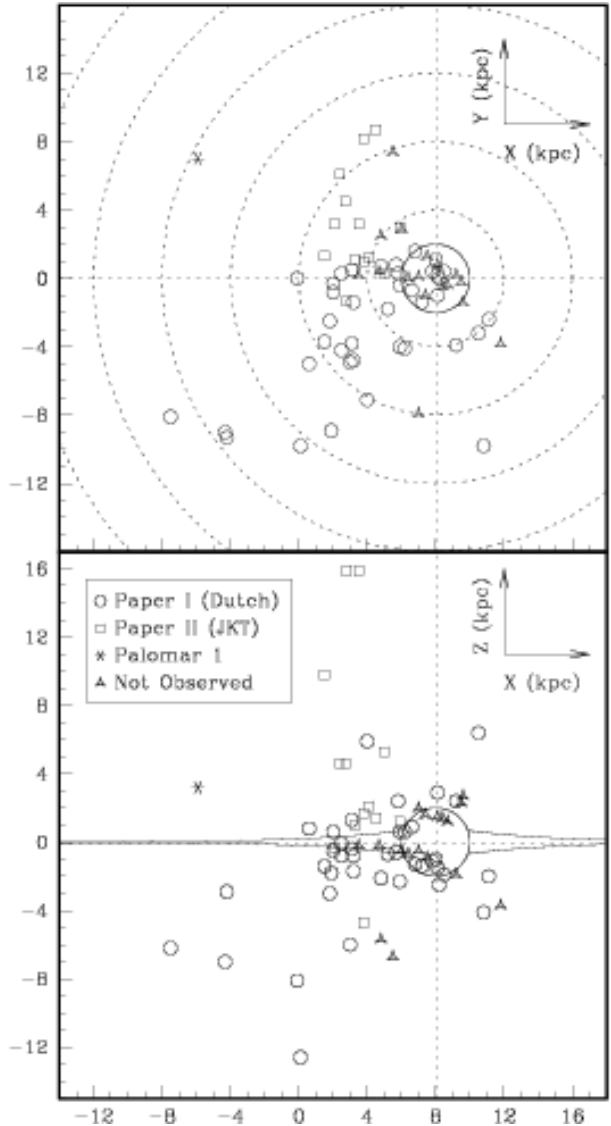,width=8cm}
\caption{
Galactocentric distribution of all the GGCs with $(m-M)_{V}\leq
16.15$~mag. In the {\it upper panel}, the projection of the GGC
location over the Galactic plane is presented. The {\it open circles}
represent the GGCs presented in Paper I, the {\it open squares} the
GGCs whose photometry and CMDs are discussed in this paper, and the
asterisk indicates the position of the GGC Pal~1 (Rosenberg et al
\cite{rosenberg98}). The clusters which are not presented in our
catalog are marked by open triangles. The {\it lower panel} shows the
location of the same GGCs in a plane perpendicular to the Galactic
plane. The Milky Way is also schematically represented.\label{f:galdist}}
\end{figure}

As discussed in Rosenberg et al. (\cite{rosenberg00}, hereafter Paper
I), the heterogeneity of the data often used in the literature for
large scale studies of the Galactic globular cluster (GGC) properties
has induced us to start a large survey of both southern and northern
GGCs by means of 1-m class telescopes, i.e. the 91cm European Southern
Observatory (ESO) / Dutch telescope and the 1m Isaac Newton Group
(ING) / Jacobus Kapteyn telescope (JKT). We were able to collect the
data for 52 of the 69 known GGCs with $(m-M)_{\rm V}\leq16.15$.
Thirty-nine objects have been observed with the Dutch Telescope and
the data have been already presented in Paper I. The images and the
photometry of the remaining 13 GGCs, observed with the JKT are
presented in this paper. A graphical representation of the spatial
distribution of our cluster sample is given in Fig.~\ref{f:galdist}

As a first exploitation of this new data base, we have conducted a GGC
relative age investigation based on the best 34 CMDs of our catalog
(Rosenberg et al. \cite{rosenberg99}, hereafter Paper III), showing
that most of the GGCs have the same age. We have also used the data
base to obtain a photometric metallicity ranking scale (Saviane et
al. \cite{saviane00}, hereafter Paper IV), based on the red giant
branch (RGB) morphology.

There are many other parameters that can be measured from an
homogeneous, well calibrated CMD data base: the horizontal branch (HB)
level, homogeneous reddening and distance scales, etc. We are
presently working on these problems. However, we believe it is now the
time to present to the community the complete data base to give to
anyone interested the opportunity to take advantage of it.

In the next section, we will describe the observations collected at
the JKT in 1997. The data reduction and calibration is presented in
Sect.~\ref{dat}, where a comparison of the calibration of the northern
and southern clusters is also discussed for three objects observed
with both telescopes.  In order to assist the reader, in
Sect.~\ref{parameters} we present the main parameters characterizing
our clusters.  Finally, the observed fields for each cluster, and the
obtained CMDs are presented and briefly discussed in Sect.~\ref{cmds}.

\section{Observations} 
\label{obs}

The data were collected on May 30-June 2 1997. For almost 2 nights we
had quite good seeing conditions (FWHM between 0.65 and 0.85
arcsec). Unfortunately, too few standards were observed during this
run.  In order to ensure an homogeneous calibration, we reobserved all
the same cluster fields (plus numerous standard fields) in June 24,
1998, a night that was photometric and with a stable seeing.

Observations were done at the 1m Jacobus Kapteyn telescope (JKT),
located at the Roque de Los Muchachos Observatory. The same
instrumentation was used in both runs: A thinned Tektronix CCD with
$1024\times1024$ pixels projecting 0.331 arcsec on the sky and
providing a field of view of 5.6 arcmin square. The detector system
was linear to $0.1\%$ over the full dynamic range of the 16 bit
analog-to-digital converter.

Three short ($\rm 10s$), two medium ($\rm 45s\div120s$) and one large
($\rm 1800s$) exposures were taken for one field of each of the
proposed objects in each filter. Also a large number of standard stars
were obtained during the second observing run. This was done in order
to have all the clusters in exactly the same system, making internal
errors negligible.

\section{Data reduction and calibration}
\label{dat}

The data were reduced following the same procedure described in Paper I.

The absolute calibration of the observations is based on a set of
standard stars of the catalog of Landolt (\cite{l92}). Eighteen
standard stars were observed; specifically, the observed fields were
PG1525, PG1633, Mark-A and PG2213. At least 4 exposures at different
airmasses were taken during the night for each standard field, making
a total of $\sim 80$ individual measures per filter.

\begin{figure}
\psfig{figure=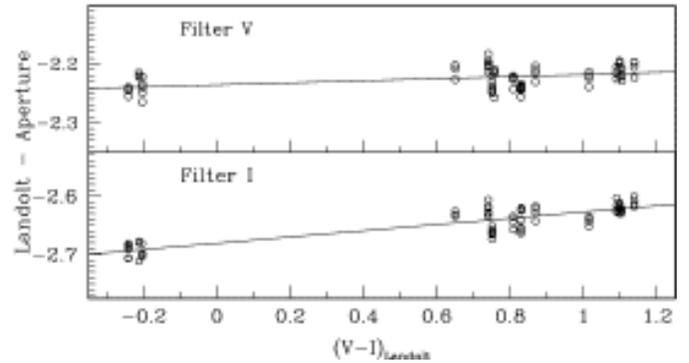,width=8.8cm}
\caption[]{
Calibration curves for the $V$ ({\it upper panel}) and $I$ ({\it lower
panel}) filters for the June $24^{\rm th}$ 1998 run.\label{f:calcurv}}
\end{figure}

Thanks to the relatively stable seeing conditions, for the aperture
photometry we used for all the standards (and the cluster fields) a 12
pixel aperture (4 arcsec).  The aperture magnitudes were normalized by
correcting for the exposure time and airmass. Using the standards
observed at different airmasses, we estimated for the extinction
coefficients in the two filters: $A_{\rm V}=0.11$ and $A_{\rm
I}=0.06$.  For the calibration curves we adopted a linear
relation. The best fitting straight lines are:\\ 
$ V-v(ap)=(0.017\pm 0.003) \times (V-I)-(2.232\pm 0.002)$; \\ 
$ I-i(ap)=(0.053\pm 0.003) \times (V-I)-(2.679\pm 0.002)$.\\
The magnitude differences vs. the standard color are plotted in
Fig.~\ref{f:calcurv}, where the solid lines represent the above equations.

\subsection{Comparison between the two catalogs}

Apart from the 13 GGCs presented in the following sections, we
re-observed three additional clusters selected from the southern
hemisphere catalog presented in Paper I.  We covered the same spatial
region, in order to test the homogeneity of the two catalogs created
using the two different telescopes.

\begin{figure}
\psfig{figure=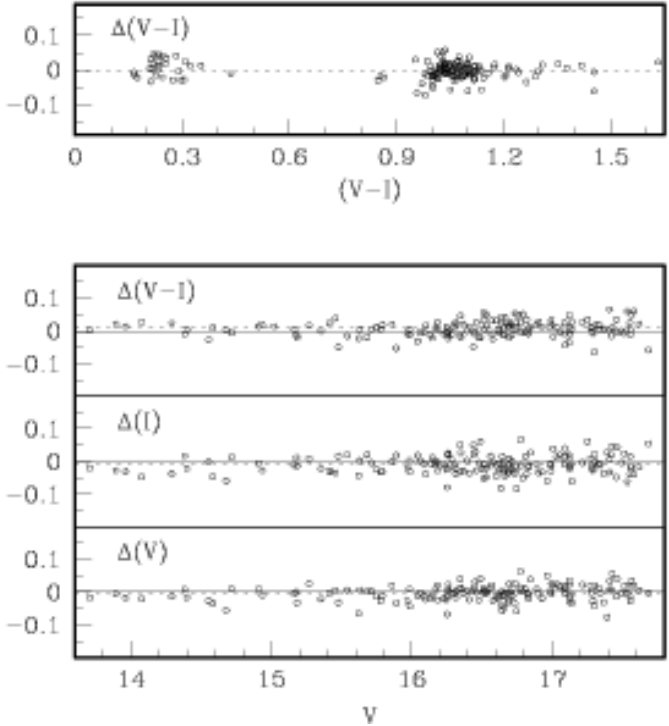,width=8.8cm}
\caption[]{
Comparison between the JKT and the ESO/Dutch telescope magnitudes for
the 170 stars in NGC~5897 with errors $\leq0.025$. The differences in
magnitude and color between the two runs are: $0.00\pm0.02$,
$-0.01\pm0.02$ and $0.01\pm0.02$ in $V$, $I$ and $V-I$,
respectively. Notice also that the slopes of the differences in color
vs.  magnitudes or color of one of the runs ({\it upper panel}), are
negligible ($\leq 0.001 \pm 0.002$), making the catalogs
photometrically homogeneous.}
\label{comparison}
\end{figure}

In Fig.~\ref{comparison} the magnitudes of 170 stars (with photometric
error smaller than 0.025~mag and $V$ magnitude brighter than 18) in
NGC~5897 from the JKT and the ESO/Dutch telescope data sets are
compared. The differences in magnitude and color between the two
observing runs are: $0.00\pm0.02$, $-0.01\pm0.02$ and $0.01\pm0.02$ in
$V$, $I$ and $V-I$, respectively.  For the other two clusters, we
have: $\Delta V^{\rm JKT}_{\rm DUT}=0.02\pm0.02$, $\Delta I^{\rm
JKT}_{\rm DUT}=0.01\pm0.02$ and $\Delta (V-I)^{\rm JKT}_{\rm
DUT}=0.01\pm0.02$ for 163 stars selected in NGC~6093, and $\Delta
V^{\rm JKT}_{\rm DUT}=0.01\pm0.02$, $\Delta I^{\rm JKT}_{\rm
DUT}=0.00\pm0.02$ and $\Delta (V-I)^{\rm JKT}_{\rm DUT}=0.01\pm0.02$
for 249 stars selected in NGC~6171 (selection criteria as in the case
of NGC~5897).  Moreover, the slopes of the color differences vs. both
the magnitude and the color are always $\leq 0.001 \pm
0.002$. Therefore, the two catalogs must be considered photometrically
homogeneous and the measures of the absolute and relative parameters
of the CMDs perfectly compatible.

\section{Parameters for the  GGC sample.}
\label{parameters}

\begin{table*}
\label{param1}
\caption[]{Identifications, positional data and metallicity
for the observed clusters.}
\begin{center}
\begin{tabular}{rlccccccccclc}
\noalign{\smallskip}
\hline
\noalign{\smallskip}
\multicolumn{1}{c}{ID} & 
\multicolumn{1}{c}{Cluster} & 
\multicolumn{1}{c}{$AR^{ \mathrm{a}}$} & 
\multicolumn{1}{c}{$DEC^{ \mathrm{b}}$} &
\multicolumn{1}{c}{${\it l}^{ \mathrm{c}}$} & 
\multicolumn{1}{c}{${\it b}^{ \mathrm{d}}$} &
\multicolumn{1}{c}{$R_{\odot}^{ \mathrm{e}}$} &
\multicolumn{1}{c}{$R_{\rm GC}^{ \mathrm{f}}$} &
\multicolumn{1}{c}{$X^{ \mathrm{g}}$} &
\multicolumn{1}{c}{$Y^{ \mathrm{h}}$} &
\multicolumn{1}{c}{$Z^{ \mathrm{i}}$} & 
\multicolumn{2}{c}{[Fe/H]} \\
\multicolumn{1}{c}{} & 
\multicolumn{1}{c}{} & 
\multicolumn{1}{c}{($^{\rm h \; m \; s}$)} & 
\multicolumn{1}{c}{($^{\rm o} \;$ \arcmin $\;$ \arcsec)} &
\multicolumn{1}{c}{$(^{\rm o})$} & 
\multicolumn{1}{c}{$(^{\rm o})$} &
\multicolumn{1}{c}{(kpc)} &
\multicolumn{1}{c}{(kpc)} &
\multicolumn{1}{c}{(kpc)} &
\multicolumn{1}{c}{(kpc)} &
\multicolumn{1}{c}{(kpc)} &
\multicolumn{1}{c}{ZW84$^{ \mathrm{j}}$} &
\multicolumn{1}{c}{CG97$^{ \mathrm{k}}$} \\

\noalign{\smallskip}
\hline 
\noalign{\smallskip}
 1 & NGC~5053 & 13 16 27.0 & +17 41 53 & 335.69 & +78.94 & 16.2 & 16.8 & +2.8 & -1.3 & +15.9 & $-2.10        $ & $-1.78      $  \\
 2 & NGC~5272 & 13 42 11.2 & +28 22 32 & 042.21 & +78.71 & 10.0 & 11.9 & +1.5 & +1.3 &  +9.8 & $-1.66^{\rm 1}$ & $-1.34^{\rm 2}$\\
 3 & NGC~5466 & 14 05 27.3 & +28 32 04 & 042.15 & +73.59 & 16.6 & 16.9 & +3.5 & +3.2 & +15.9 & $-2.22^{\rm 1}$ & $    -$        \\
 4 & NGC~5904 & 15 18 33.8 & +02 04 58 & 003.86 & +46.80 &  7.3 &  6.1 & +5.0 & +0.3 &  +5.3 & $-1.38        $ & $-1.12$        \\
\hline 
 5 & NGC~6205 & 16 41 41.5 & +36 27 37 & 059.01 & +40.91 &  7.0 &  8.3 & +2.7 & +4.5 &  +4.6 & $-1.63        $ & $-1.33$        \\
 6 & NGC~6218 & 16 47 14.5 & -01 56 52 & 015.72 & +26.31 &  4.7 &  4.6 & +4.1 & +1.2 &  +2.1 & $-1.40        $ & $-1.14$        \\
 7 & NGC~6254 & 16 57 08.9 & -04 05 58 & 015.14 & +23.08 &  4.3 &  4.6 & +3.8 & +1.0 &  +1.7 & $-1.55        $ & $-1.25$        \\
 8 & NGC~6341 & 17 17 07.3 & +43 08 11 & 068.34 & +34.86 &  8.1 &  9.5 & +2.4 & +6.1 &  +4.6 & $-2.24^{\rm 1}$ & $-2.16^{\rm 2}$\\
\hline 
 9 & NGC~6366 & 17 27 44.3 & -05 04 36 & 018.41 & +16.04 &  3.6 &  4.9 & +3.3 & +1.1 &  +1.0 & $-0.58        $ & $-0.73$        \\
10 & NGC~6535 & 18 03 50.7 & -00 17 49 & 027.18 & +10.44 &  6.8 &  3.9 & +5.9 & +3.0 &  +1.2 & $-1.78        $ & $-1.51$        \\
11 & NGC~6779 & 19 16 35.5 & +30 11 05 & 062.66 & +08.34 &  9.9 &  9.5 & +4.5 & +8.7 &  +1.4 & $-1.94^{\rm 1}$ & $    -$        \\
12 & NGC~6838 & 19 53 46.1 & +18 46 42 & 056.74 & -04.56 &  3.8 &  6.7 & +2.1 & +3.2 &  -0.3 & $-0.58        $ & $-0.73$        \\
\hline 
13 & NGC~7078 & 21 29 58.3 & +12 10 01 & 065.01 & -27.31 & 10.2 & 10.3 & +3.8 & +8.2 &  -4.7 & $-2.13        $ & $-2.02$        \\
\noalign{\smallskip}
\hline
\end{tabular}
\\
In the following cases, the [Fe/H] values were taken directly from
($^{\rm 1}$) ZW84
and ($^{\rm 2}$) CG97.
\end{center}
\begin{tabular}{lrl}
$^{\mathrm{a}}$ Right Ascension (2000)&
Sun-Centered coordinates: & 
$^{\mathrm{g}}$ X: Toward the Galactic Center\\
$^{\mathrm{b}}$ Declination (2000) &
 &
$^{\mathrm{h}}$ Y: in direction of Galactic rotation \\
$^{\mathrm{c}}$ Galactic Longitude &
 &
$^{\mathrm{i}}$ Z: Towards North Galactic Plane \\
$^{\mathrm{d}}$ Galactic Latitude &
 & 
 \\
$^{\mathrm{e}}$ Heliocentric Distance &
[Fe/H] (From Rutledge et al. 1997):&
$^{\mathrm{j}}$ in the ZW84 scale\\
$^{\mathrm{f}}$ Galactocentric Distance &
 & 
$^{\mathrm{k}}$ in the CG97 scale\\
\end{tabular}

\end{table*}


\begin{table*}
\begin{center}
\caption[]{Photometric Parameters.}
\label{param2}
\begin{tabular}{rlcccrrccccrrc}
\noalign{\smallskip}
\hline
\noalign{\smallskip}

\multicolumn{1}{r}{ID} & 
\multicolumn{1}{c}{Cluster} & 
\multicolumn{1}{c}{${\rm E(B-V)^{ \mathrm{a}}}$} & 
\multicolumn{1}{c}{$V_{\rm HB} ^{ \mathrm{b}}$} & 
\multicolumn{1}{c}{${\rm (m-M)_{\rm V}^{ \mathrm{c}}}$} & 
\multicolumn{1}{r}{${\rm V_t}^{ \mathrm{d}}$} & 
\multicolumn{1}{r}{${\rm Mv_t}^{ \mathrm{e}}$} &
\multicolumn{1}{c}{\rm $U-B^{ \mathrm{f}}$} & 
\multicolumn{1}{c}{\rm $B-V^{ \mathrm{f}}$} & 
\multicolumn{1}{c}{\rm $V-R^{ \mathrm{f}}$} & 
\multicolumn{1}{c}{\rm $V-I^{ \mathrm{f}}$} & 
\multicolumn{1}{r}{$S_{\rm RR} ^{ \mathrm{g}}$} & 
\multicolumn{1}{r}{$HBR^{ \mathrm{h}}$} \\

\noalign{\smallskip}
\hline 
\noalign{\smallskip}
 1 & NGC~5053 & 0.03 & 16.70 & 16.14 &  9.47 & -6.67 & 0.03 & 0.65 & -    & 0.90 & 21.4  & 0.52 \\
 2 & NGC~5272 & 0.01 & 15.58 & 15.04 &  6.19 & -8.85 & 0.09 & 0.69 & -    & 0.93 & 75.3  & 0.08 \\
 3 & NGC~5466 & 0.00 & 16.60 & 16.10 &  9.04 & -7.06 & 0.00 & 0.67 & -    & 0.82 & 41.9  & 0.58 \\
 4 & NGC~5904 & 0.03 & 15.00 & 14.41 &  5.65 & -8.76 & 0.17 & 0.72 & 0.45 & 0.95 & 38.4  & 0.31 \\
\hline 
 5 & NGC~6205 & 0.02 & 14.95 & 14.28 &  5.78 & -8.50 & 0.02 & 0.68 & -    & 0.86 &  2.0  & 0.97 \\
 6 & NGC~6218 & 0.19 & 14.70 & 13.97 &  6.70 & -7.27 & 0.20 & 0.83 & 0.56 & 1.14 &  0.0  & 0.97 \\
 7 & NGC~6254 & 0.28 & 15.05 & 14.03 &  6.60 & -7.43 & 0.23 & 0.90 & 0.53 & 1.21 &  1.1  & 0.98 \\
 8 & NGC~6341 & 0.02 & 15.20 & 14.59 &  6.44 & -8.15 & 0.01 & 0.63 & -    & 0.88 & 13.7  & 0.91 \\
\hline 
 9 & NGC~6366 & 0.69 & 15.65 & 14.92 &  9.20 & -5.72 & 0.97 & 1.44 & -    & -    &  5.1  &-0.97 \\
10 & NGC~6535 & 0.32 & 15.80 & 15.15 & 10.47 & -4.68 & 0.29 & 0.94 & -    & -    &  0.0  & 1.00 \\
11 & NGC~6779 & 0.20 & 16.30 & 15.60 &  8.27 & -7.33 & 0.15 & 0.86 & -    & 1.16 &  2.3  & 0.98 \\
12 & NGC~6838 & 0.25 & 14.50 & 13.70 &  8.19 & -5.51 & 0.63 & 1.09 & 0.63 & 1.36 &  0.0  &-1.00 \\
\hline 
13 & NGC~7078 & 0.09 & 15.90 & 15.31 &  6.20 & -9.11 & 0.06 & 0.68 & -    & 0.85 & 25.4  & 0.67 \\
\noalign{\smallskip}
\hline
\end{tabular}
\end{center}

\begin{tabular}{llll}
$^{\mathrm{a}}$ Foreground reddening &
$^{\mathrm{e}}$ Absolute visual magnitude \\
$^{\mathrm{b}}$ HB Level (our measures)&
$^{\mathrm{f}}$ Integrated color indices \\
$^{\mathrm{c}}$ Apparent visual distance modulus &
$^{\mathrm{g}}$ Specific frequency of RR Lyrae variables \\
$^{\mathrm{d}}$ Integrated $V$ mag. of clusters &
$^{\mathrm{h}}$ HB ratio: $HBR=(B-R)/(B+V+R)$ \\
\end{tabular}
\end{table*}


\begin{table*}
\begin{center}
\caption[]{Kinematics, and Structural Parameters}
\label{param3}
\begin{tabular}{rlrrccccccc}
\noalign{\smallskip}
\hline
\noalign{\smallskip}
ID & 
Cluster & 
$V_{\rm r}^{ \mathrm{a}}$ & 
$V_{\rm LSR}^{ \mathrm{b}}$ & 
$c^{ \mathrm{c}}$ & 
$r_{\rm c}^{ \mathrm{d}}$ & 
$r_{\rm h}^{ \mathrm{e}}$ & 
$lg(t_{\rm c})^{ \mathrm{f}}$ & 
$lg(t_{\rm h})^{ \mathrm{g}}$ & 
$\mu_{\rm V}^{ \mathrm{h}}$ & 
$\rho_{\rm 0}^{ \mathrm{i}}$ \\
\noalign{\smallskip}
\hline 
\noalign{\smallskip}
 1 & NGC~5053 &  $ +44.0\pm0.4$ & $ +51.8$ &  0.82 & 2.25 & 3.50 & 9.87 &  9.59 &  22.19 & 0.47 \\
 2 & NGC~5272 &  $-148.6\pm0.4$ & $-138.0$ &  1.85 & 0.50 & 1.12 & 8.78 &  9.02 &  16.34 & 3.56 \\
 3 & NGC~5466 &  $+107.7\pm0.3$ & $ 119.7$ &  1.43 & 1.96 & 2.25 & 9.75 &  9.37 &  21.28 & 0.78 \\
 4 & NGC~5904 &  $ +52.1\pm0.5$ & $ +65.2$ &  1.87 & 0.40 & 2.11 & 8.30 &  9.13 &  16.05 & 3.94 \\
\hline 
 5 & NGC~6205 &  $-246.6\pm0.9$ & $-228.2$ &  1.49 & 0.88 & 1.49 & 8.94 &  8.96 &  16.80 & 3.32 \\
 6 & NGC~6218 &  $ -42.1\pm0.6$ & $ -26.3$ &  1.38 & 0.66 & 2.16 & 8.23 &  8.86 &  18.17 & 3.27 \\
 7 & NGC~6254 &  $ +75.8\pm1.0$ & $ +91.5$ &  1.40 & 0.86 & 1.81 & 8.38 &  8.56 &  17.69 & 3.51 \\
 8 & NGC~6341 &  $-121.6\pm1.5$ & $-103.0$ &  1.81 & 0.23 & 1.09 & 7.84 &  8.56 &  15.58 & 4.30 \\
\hline 
 9 & NGC~6366 &  $-122.3\pm0.5$ & $-106.3$ &  0.92 & 1.83 & 2.63 & 8.77 &  8.50 &  21.24 & 2.39 \\
10 & NGC~6535 &  $-215.1\pm0.5$ & $-198.0$ &  1.30 & 0.42 & 0.77 & 7.47 &  8.05 &  20.22 & 2.66 \\
11 & NGC~6779 &  $-135.7\pm0.8$ & $-116.9$ &  1.37 & 0.37 & 1.16 & 8.27 &  8.70 &  18.06 & 3.27 \\
12 & NGC~6838 &  $ -22.9\pm0.2$ & $  -5.5$ &  1.15 & 0.63 & 1.65 & 7.62 &  8.10 &  19.22 & 3.06 \\
\hline 
13 & NGC~7078 &  $-107.5\pm0.3$ & $ -95.0$ &  2.50 & 0.07 & 1.06 & 7.10 &  9.01 &  14.21 & 5.37 \\
\noalign{\smallskip}
\hline
\end{tabular}
\smallskip

\begin{tabular}{lll}

$^{\mathrm{a}}$Heliocentric radial velocity &
$^{\mathrm{d}}$The core radii &
$^{\mathrm{g}}$Log. of core relaxation time at $r_{\rm h}$ \\
$^{\mathrm{b}}$Radial velocity relative to the(LSR) &
$^{\mathrm{e}}$The core median radii &
$^{\mathrm{h}}$Central surface brightness \\
$^{\mathrm{c}}$Concentration parameter $[c=log(r_{\rm t}/r_{\rm c})]$ &
$^{\mathrm{f}}$Log. of relaxation time in years &
$^{\mathrm{i}}$Log. of central luminosity density \\
\end{tabular}
\end{center}
\end{table*}

In order to assist the reader, we present in Tables 1\EXparamUno,
2\EXparamDue\ and 3\EXparamTre\ the basic parameters available for our
GGC sample\footnote {Unless otherwise stated, the data presented in
these tables are taken from the McMaster catalog described by Harris
(\cite{harris96}).}.

In Table~1\EXparamUno\ we give the coordinates, the position, and the
metallicity of the clusters: right ascension and declination (epoch
J2000, cols. 3 and 4); Galactic longitude and latitude (cols. 5 and
6); Heliocentric (col. 7) and Galactocentric (col. 8) distances
(assuming $R_{\sun}$=8.0 kpc); spatial components (X,Y,Z) (cols 9, 10
and 11) in the Sun-centered coordinate system (X pointing toward the
Galactic center, Y in direction of Galactic rotation, Z toward North
Galactic Pole) and, finally, the metallicity given in Rutledge et
al. (\cite{rutledge97}), on both the Zinn \& West (\cite{zinnwest84})
and Carretta \& Gratton (\cite{carretagratton97}) scales.

In Table~2\EXparamDue, the photometric parameters are given.  Column 3
lists the foreground reddening; column 4, the $V$ magnitude level of
the horizontal branch; column 5, the apparent visual distance modulus;
integrated $V$ magnitudes of the clusters are given in column 6;
column 7 gives the absolute visual magnitude. Columns 8 to 11 give the
integrated color indices (uncorrected for reddening). Column 12 gives
the specific frequency of RR Lyrae variables, while column 13 list the
horizontal-branch morphological parameter (Lee \cite{lee90}).

In Table~3\EXparamTre, we present the kinematical and structural
parameters for the observed clusters. Column 3 gives the heliocentric
radial velocity (km/s) with the observational (internal) uncertainty;
column 4, the radial velocity relative to the local standard of rest;
column 5, the concentration parameter ($c = \log (r_{\rm t}/r_{\rm
c})$); a 'c' denotes a core-collapsed cluster; columns 6 and 7, the
core and the half mass radii in arcmin; column 8, the logarithm of the
core relaxation time, in years; and column 9 the logarithm of the
relaxation time at the half mass radius.  Column 10, the central
surface brightness in $V$; and column 11, the logarithm of central
luminosity density (Solar luminosities per cubic parsec).


\section{The Color-Magnitude Diagrams}
\label{cmds}

In this section the $V$ vs. ($V-I$) CMDs for the 13 GGCs are presented.

The same color and magnitude scales have been used in plotting the
CMDs, so that differential measurements can be done directly using the
plots.  The adopted scale is the same used in previous Paper I. Two
dot sizes have been used: the larger ones correspond to the better
measured stars, normally selected on the basis of their photometric
error ($\leq0.1$) and sharpness parameter.  In some exceptional cases,
a selection on the radial distance from the cluster center is also
done, in order to make more evident the cluster CMD over the field
stars, or to show differential reddening effects.  The smaller size
dots show all the measured stars with errors (as calculated by
DAOPHOT) smaller than 0.15 mag.

The images of the fields are oriented with North at the top and East
on the left side. Each field covers $5 \farcm 6$ square. The same
spatial scale has been used in all the cluster images.

In the next subsections, we briefly present the single CMDs and
clusters, and give some references to the best existing CMDs.  This
does not pretend to be a complete bibliographical catalog: a large
number of CMDs are available in the literature for many of the
clusters of this survey; we will concentrate just on the best CCD
photometric works.  The tables with the position and photometry of the
measured stars will be available on-line at the IAC
(http://www.iac.es/proyect/poblestelares) and Padova
(http://menhir.pd.astro.it/).

\begin{figure*}
\begin{tabular}{cc}
\raisebox{-2cm}{\psfig{figure=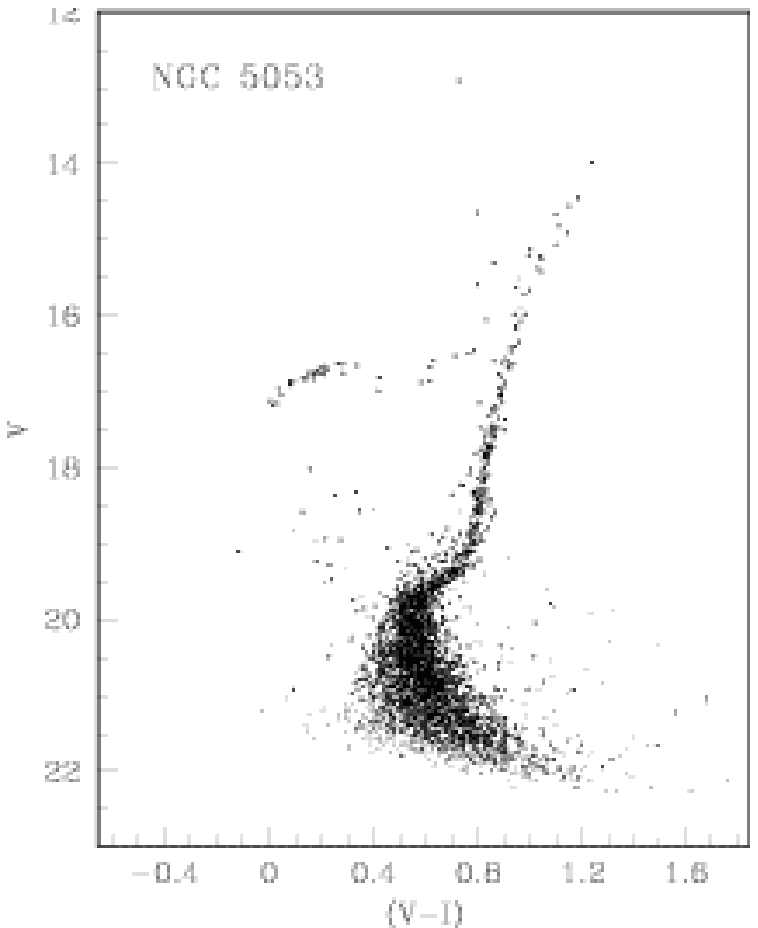,width=8.8cm}} &
\fbox{\psfig{figure=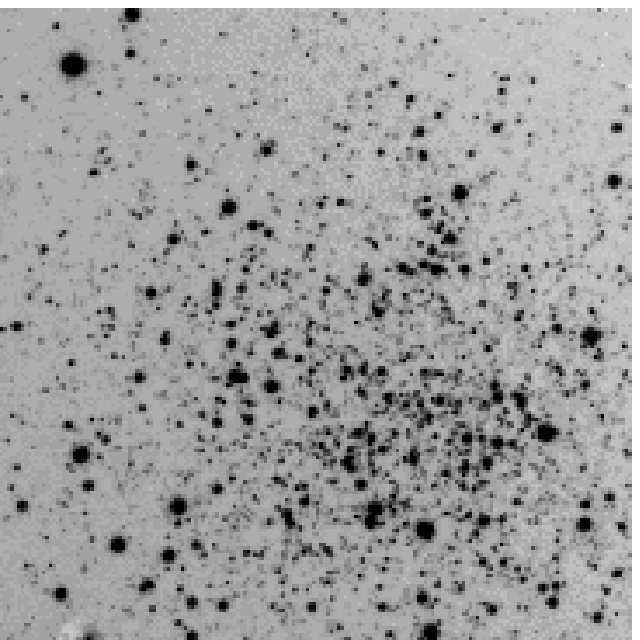,width=6cm}}
\end{tabular}
\caption[]{CMD and covered field for NGC~5053}
\label{ngc5053}
\end{figure*}

\newpage
\paragraph {\bf NGC~5053.} (Fig.~\ref{ngc5053})

NGC~5053 is a low concentration cluster, and, as all the sparse
clusters, it has a small central velocity dispersion and central mass
density.  It is one of the clusters farthest from the Galactic center
in our sample. NGC~5053 is one of the most metal-poor clusters in our
Galaxy (Sarajedini and Milone \cite{sarajedini95}).

There are several CMD studies for NGC~5053 in the literature. Nemec \&
Cohen (\cite{nemec89}) presented the first CCD CMD in the Thuan \&
Gunn (\cite{thuan76}) $g$ and $r$ filters, reaching $g\simeq23$.
Their CMD is one magnitude deeper than ours, but the stellar
distribution and number of stars above the $\rm 22^{\rm nd}$ magnitude
is almost the same in both diagrams. The upper part of their RGB
(above the HB) is saturated.  Heasley \& Christian (\cite{heasley91a})
present $B$ and $V$ photometry extended to $V\simeq22$.  Their CMD is
poorly populated, presenting only a few stars in the HB region. Their
upper RGB is also truncated at $V\sim15.5$.  In the same year, Fahlman
et al. (\cite{fahlman91}) present a study of the stellar content and
structure of this cluster, including a $B$ and $V$ CMD that reaches
$V=24$. They are mainly interested in the stellar content and
structure, being most of the data obtained just for the $V$ band, and
only one field (field \#2) in both, $V$ and $B$ bands. The
corresponding CMD is deep but poorly populated, presenting $\sim 6$
stars in their HB.

More recently, Sarajedini and Milone (\cite{sarajedini95}) present
$B$, $V$, and $I$ photometry for the upper part of the CMD (above the
cluster's TO), making a good sampling ($25\%$ larger than ours) of the
evolved cluster stars.

We present a photometry ($\sim5300$ stars, seeing of $\sim1.1\arcsec$) that
covers the cluster from the brighter RGB stars down to the $\rm
22^{\rm nd}$ magnitude. All the CMD sequences are well defined,
including a blue straggler sequence. NGC~5053 has a BHB with a few
RR-Lyrae and also a few stars in the red side of the HB (evolved HB
stars?).

\begin{figure*}
\begin{tabular}{cc}
\raisebox{-2cm}{\psfig{figure=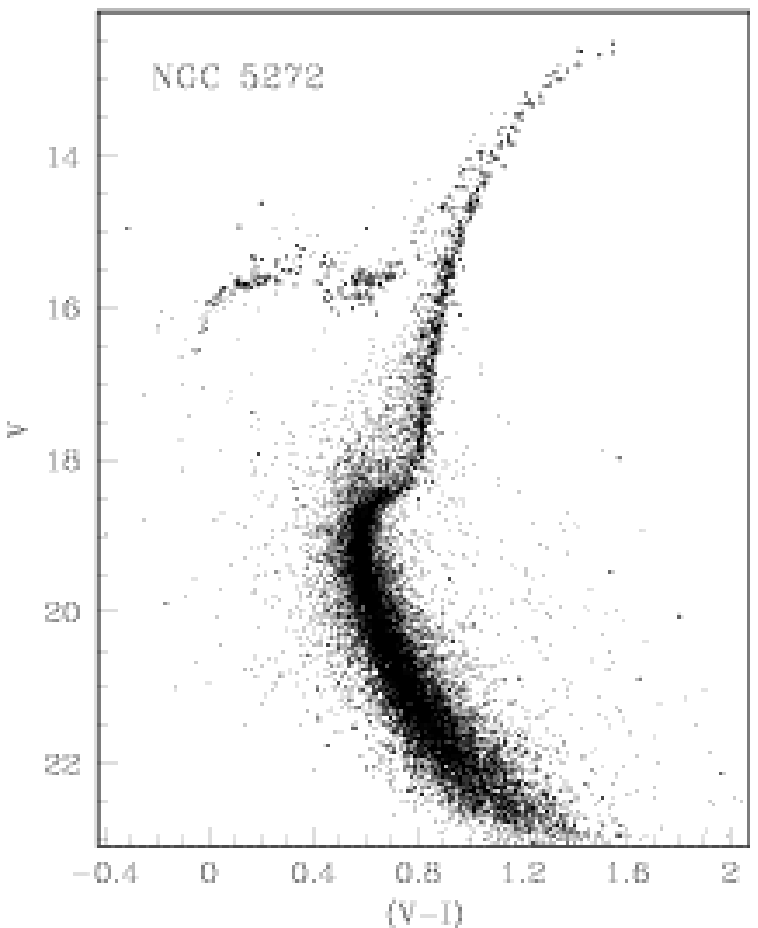,width=8.8cm}} &
\fbox{\psfig{figure=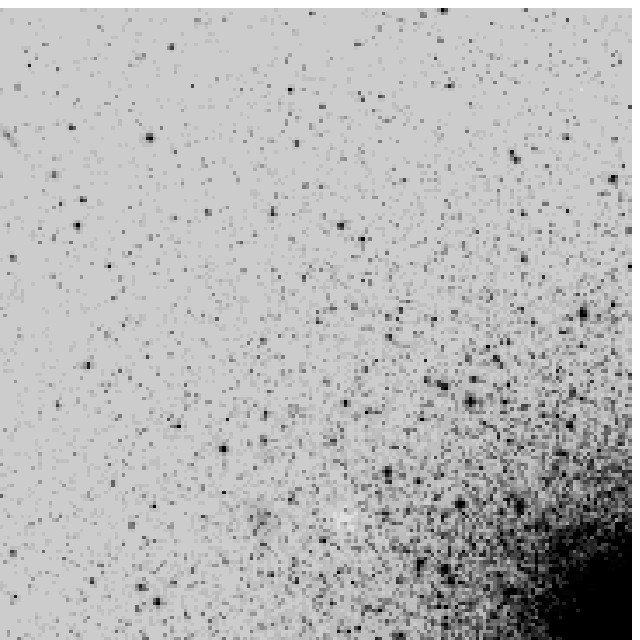,width=6cm}}
\end{tabular}
\caption[]{CMD and covered field for NGC~5272 (M~3)}
\label{ngc5272}
\end{figure*}

\paragraph {\bf NGC~5272 (M~3).} (Fig.~\ref{ngc5272})

The northern ``standard couple'' of clusters affected by the second
parameter effect is represented by M~3 and M~13. M~3 has a well
populated HB, both on the red and blue sides, while the M~13 HB is
populated only on the blue side.  This is a typical example where the
age cannot be advocated as the (unique) responsible of the differences
in the HB morphology (Paper III, Johnson \& Bolte \cite{johnson98},
Davidge \& Courteau \cite{davidge99}).

Very recently, M3 has been the subject of many studies. Laget et
al. (\cite{laget98}) present $UV$, $U$, $V$ and $I$ photometry of the
central part of M~3 from HST data. Their CMDs reach $V\sim20$, and is
extremely well defined. Their stellar sample is smaller than our one,
but clearly shows the mean regions of the CMD. Its overall structure
is very similar to our ``selected'' diagram. Kaluzny et
al. (\cite{kaluzny98}) look for variable stars in the $B$ and $V$
bands. Their CMD has a limiting magnitude $V\sim21$, is less populated
(probably a factor of two, based on the number of BHB stars), since
the MS stars are very sparse.  An excellent CMD in the $V$ and $I$
filters is presented by Johnson \& Bolte (\cite{johnson98}). Our field
does not match that covered by them. However, the fiducial line
representing their CMD well matches our diagram, with a small zero
point difference (of the order of 0.02 magnitudes in V) in the RGB: a
difference within the uncertainties of our absolute
calibration. Recently, Davidge \& Courteau (\cite{davidge99})
published $JHK$ data for four clusters, including M~3.  Their CMD
extends from the brightest stars down to the (sparse) MS.  A large
sample of M3 stars ($\sim37.000$) has been measured in the $V$ and $I$
bands by Rood et al. (\cite{rood99}), combining groundbased and HST
data extending from the cluster center to the outer radius.  This CMD
is very clean as the high HST resolution limits crowding effects in
the cluster center. Notice that the RR-Lyrae stars were identified and
removed by the authors.

Our photometry covers the cluster from the tip of the RGB to the $\rm
23^{\rm rd}$ magnitude ($\sim18750$ stars), under quite good seeing
conditions. The HB bimodality is clearly visible. Stars spread above
the MS turn-off and blue-ward from the RGB are at small distances from
the cluster center. If these stars were not plotted, the CMD would
become very well defined (larger dots). Moreover, a large number of
RR-Lyrae stars are present in our diagram, and our dispersion in color
(see the MS, for example) is smaller than that present on the Rood et
al (\cite{rood99}) diagram.

\begin{figure*}
\begin{tabular}{cc}
\raisebox{-2cm}{\psfig{figure=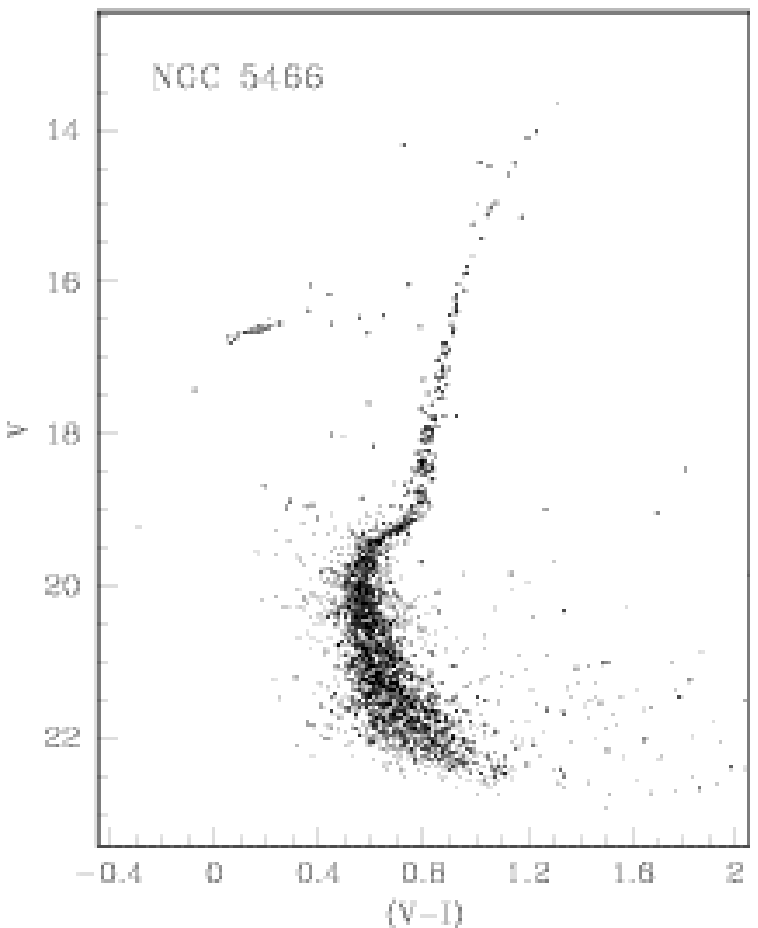,width=8.8cm}} &
\fbox{\psfig{figure=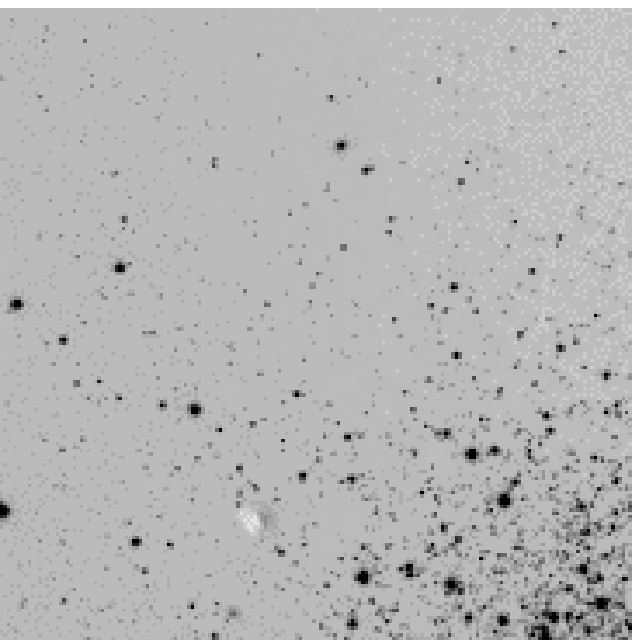,width=6cm}}
\end{tabular}
\caption[]{CMD and covered field for NGC~5466}
\label{ngc5466}
\end{figure*}

\paragraph {\bf NGC~5466.} (Fig.~\ref{ngc5466})

NGC~5466 has one of the lowest central densities among the GGCs. Previous
CMDs are not very recent; Nemec \& Harris (\cite{nemec87}) present
photographic and CCD data, with the CCD CMD extending from $V\sim21$, to
just below the HB (not present). We have not found more recent CMDs for
this cluster.

We present a CMD from the RGB tip down to $V\sim22.5$ covering
$\sim5000$ stars. NGC~5466 resembles in many aspects NGC~5053,
including the metal content, and also their CMDs look very similar.

\begin{figure*}
\begin{tabular}{cc}
\raisebox{-2cm}{\psfig{figure=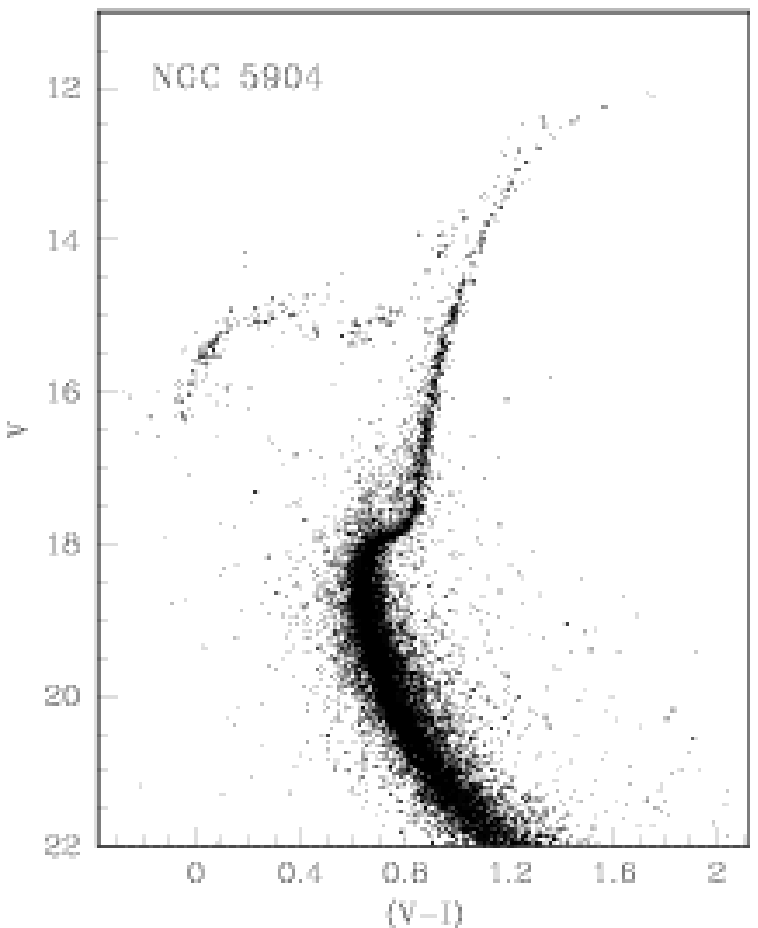,width=8.8cm}} &
\fbox{\psfig{figure=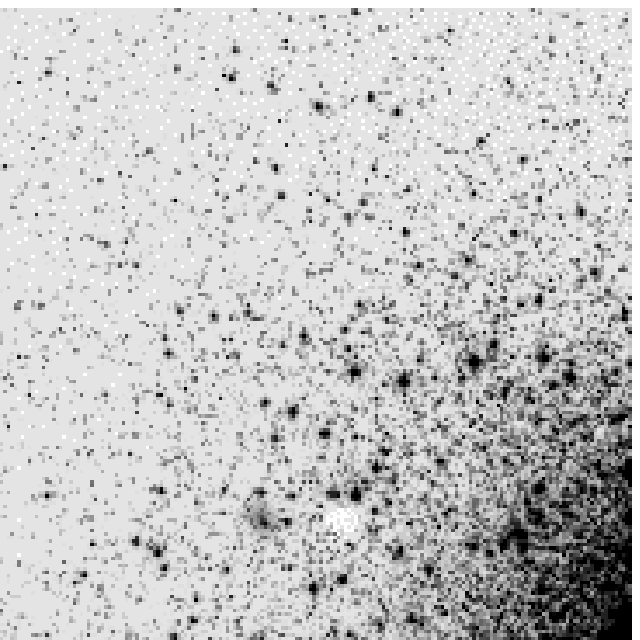,width=6cm}}
\end{tabular}
\caption[]{CMD and covered field for NGC~5904 (M~5)}
\label{ngc5904}
\end{figure*}

\paragraph {\bf NGC~5904 (M~5).} (Fig.~\ref{ngc5904})

The globular cluster M~5 harbors one of the richest collection of RR-Lyrae
stars in the Galaxy. It also hosts one of the only two known dwarf novae in
GGCs.

The first CCD CMD for this cluster is published by Richer \& Fahlman
(\cite{richer87}), who present deep $U$, $B$, and $V$ photometry. They
give a well defined diagram, but poorly populated on the RGB and the
HB. More recently, Sandquist et al. (\cite{sandquist96}) present $B$,
$V$, and $I$ photometry for more than 20.000 stars in M5, and an
excellent CMD extended down to $V\sim22$.  The latest ground-based
study is in Johnson \& Bolte (\cite{johnson98}), who presented very
good $V$ and $I$ photometry for this cluster. They compare their
photometric calibration with that of Sandquist et
al. (\cite{sandquist96}), with an uncomfortable trend with magnitude
and an offset that increases with decreasing brightness. Recently, HST
data have been published by Drissen \& Shara (\cite{drissen98}), who
studied the stellar population and the variable stars in the core of
the cluster.  Again, we compared our photometry with the fiducials
from Johnson \& Bolte (\cite{johnson98}), finding a good agreement
within the errors.

Also NGC~5904 has been observed during quite good seeing
conditions. The CMD extends by $\ge4$ magnitudes below the TO, and
includes $\sim18300$ stars. All the CMD branches are well defined. In
particular, note the perfect distinction between the AGB and the RGB,
quite rare even in recent CMDs also for other clusters, and the
extended blue straggler sequence.

\begin{figure*}
\begin{tabular}{cc}
\raisebox{-2cm}{\psfig{figure=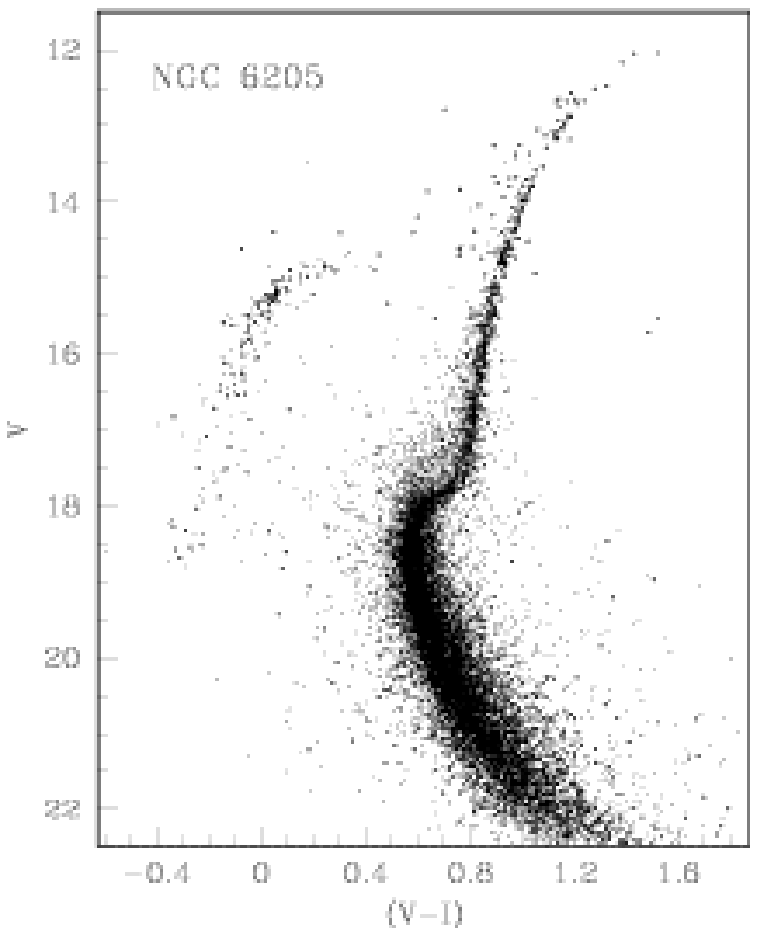,width=8.8cm}} &
\fbox{\psfig{figure=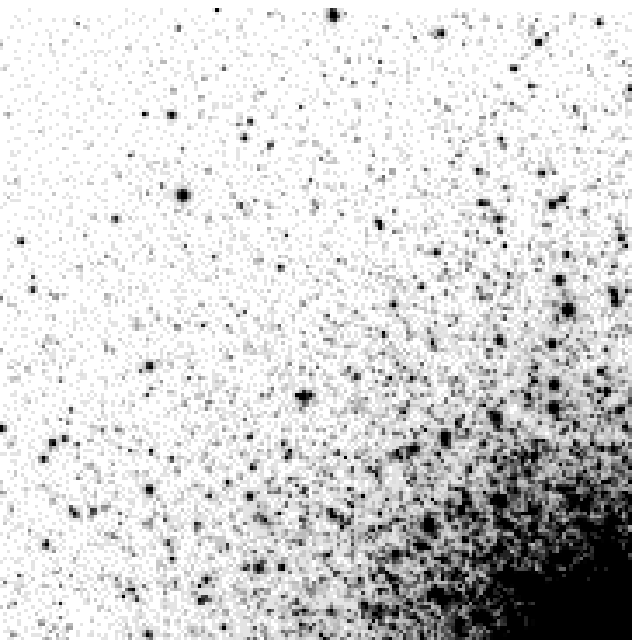,width=6cm}}
\end{tabular}
\caption[]{CMD and covered field for NGC~6205 (M~13)}
\label{ngc6205}
\end{figure*}

\paragraph {\bf NGC~6205 (M~13).} (Fig.~\ref{ngc6205})

As already mentioned, M~13 and M~3 are a classical second parameter
couple. M~13 has only a BHB, still, in several recent studies, it is
found to be coeval with M~3 (e.g. Paper III). In our CMD it shows a
sparse EBHB that arrives to the MS TO magnitude.

Previous studies include Richer \& Fahlman (\cite{richer86}) who
obtained $U$, $B$, and $V$ CCD photometry for this cluster from
$\sim2$ magnitudes above the TO to $V$ $\sim23$.  Their CMD is very
well defined, but for a very small sample of stars. Moreover, just 6
RGB stars under the HB are present.  More recently, CMDs are presented
again by Johnson \& Bolte (\cite{johnson98}), with their ($V,I$) CMD
and our one perfectly overlapping within the errors.  Paltrinieri et
al. \cite{pal98} present $B,V$ photometry for $\sim5500$ stars from
the RGB tip to about 2 magnitudes below the MS TO. Their photometry is
more sparse in the SGB and specially in the MS.  Davidge \& Courteau
(\cite{davidge99}) have published a $JHK$ photometry. They cover
basically the more evolved branches of the diagram.

\begin{figure*}
\begin{tabular}{cc}
\raisebox{-2cm}{\psfig{figure=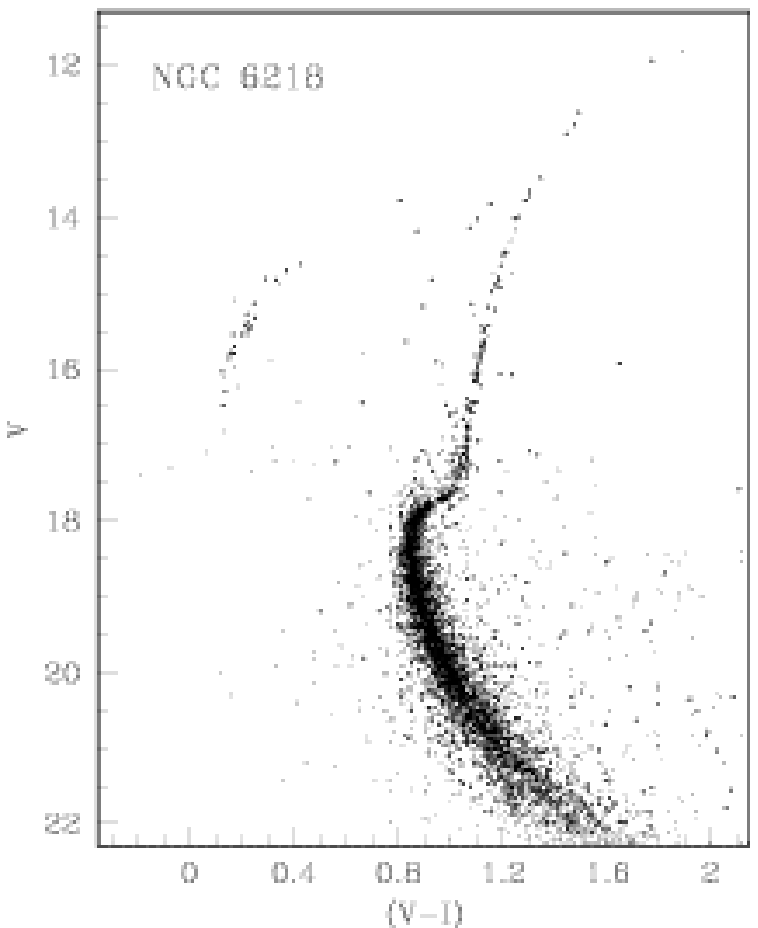,width=8.8cm}} &
\fbox{\psfig{figure=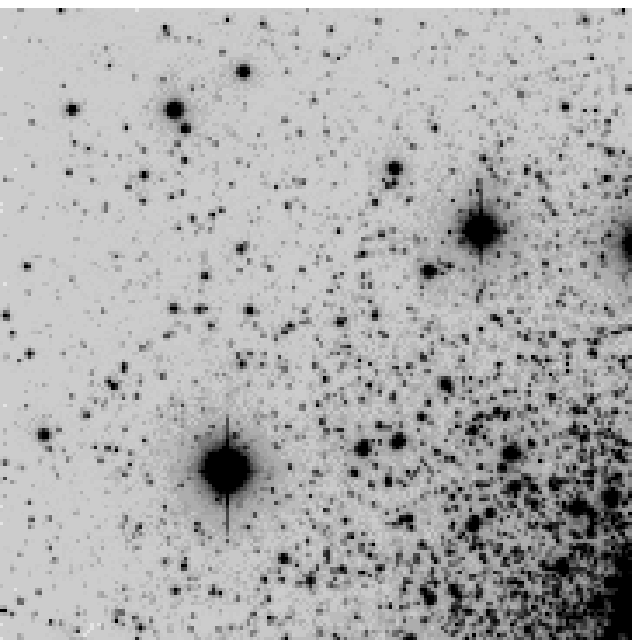,width=6cm}}
\end{tabular}
\caption[]{CMD and covered field for NGC~6218 (M~12)}
\label{ngc6218}
\end{figure*}

\paragraph {\bf NGC~6218 (M~12).} (Fig.~\ref{ngc6218})

The only two CCD studies that we have found in the literature for this
cluster are those of Sato et al. (\cite{sato89}), who presented $UBV$
data for the MS and SGB region in a poorly populated CMD, and Brocato
et al. (\cite{brocato96}), who present a sparse $B$ and $V$ CMD from
the tip of the RGB to a few magnitudes below the cluster TO.

Our CMD is not very populated ($\sim7100$ stars measured), but the
main lines of the RGB, BHB, SGB and MS are very well defined. It is
noticeable that, despite its intermediate metallicity, this cluster
shows only a BHB, resembling in some way M13.

\begin{figure*}
\begin{tabular}{cc}
\raisebox{-2cm}{\psfig{figure=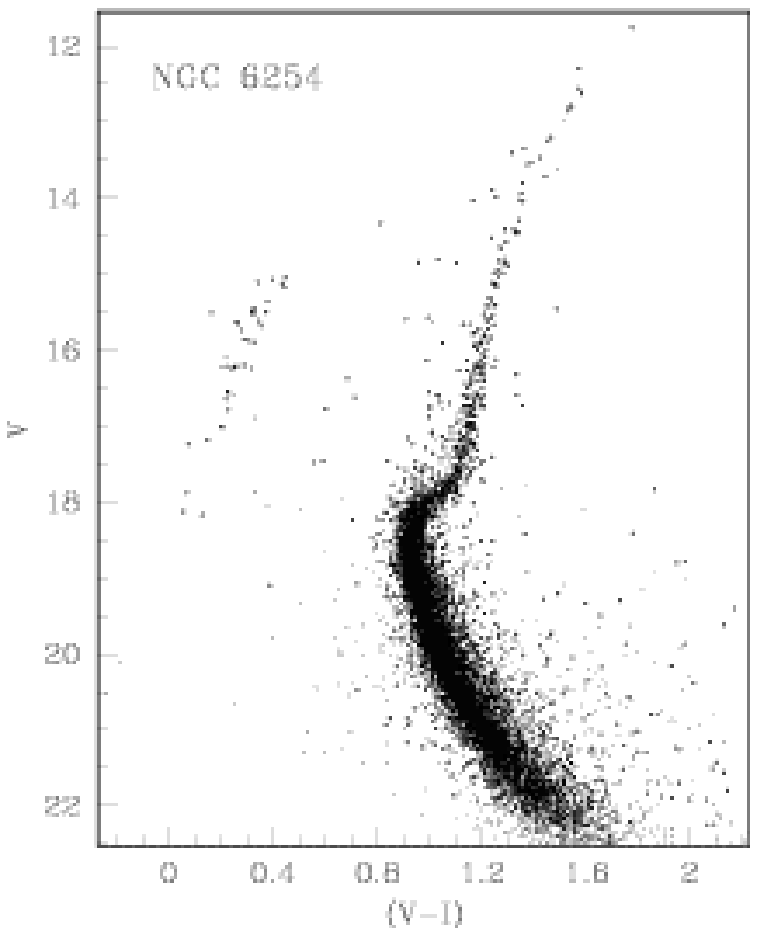,width=8.8cm}} &
\fbox{\psfig{figure=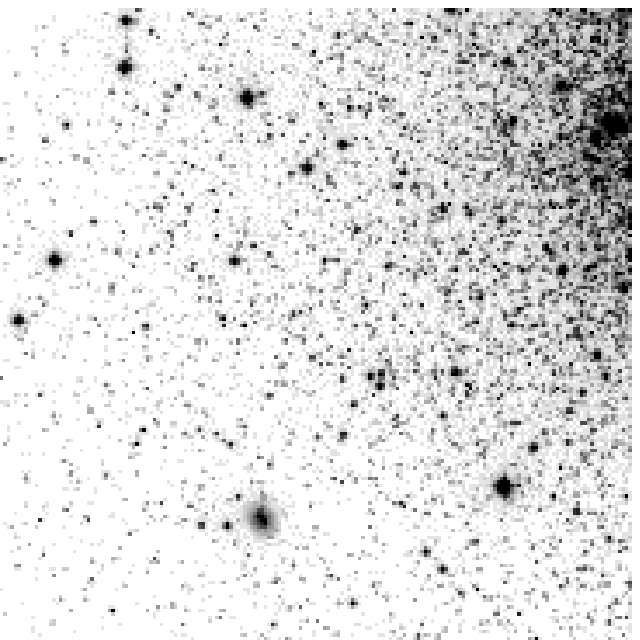,width=6cm}}
\end{tabular}
\caption[]{CMD and covered field for NGC~6254 (M~10)}
\label{ngc6254}
\end{figure*}

\paragraph {\bf NGC~6254 (M~10).} (Fig.~\ref{ngc6254})

Hurley et al. (\cite{hurley89}) present the first CCD CMD of this
cluster in the $B$ and $V$ bands, covering from the RGB (poorly
populated) to  $V\sim21.5$. However, already in this diagram the
remarkable EBHB is clearly visible. Recently, Piotto \& Zoccali
(\cite{piotto99}) present HST data for the MS of this clusters and
analyze the cluster luminosity function, using the present data for
the evolved part of the CMD.

Our CMD is well defined and extends for more than 4 magnitudes below
the TO, covering a total of $\sim13000$ stars. The cluster has only a
BHB, and we confirm that it is extended. In many respects, the CMD of
M10 resembles that of M13.

\begin{figure*}
\begin{tabular}{cc}
\raisebox{-2cm}{\psfig{figure=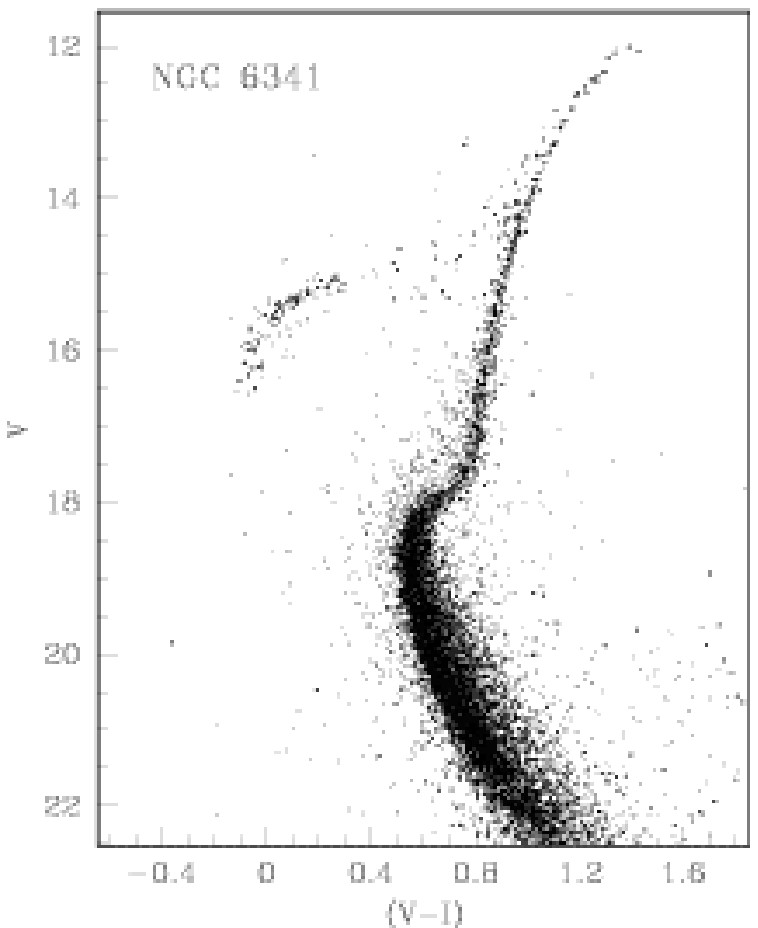,width=8.8cm}} &
\fbox{\psfig{figure=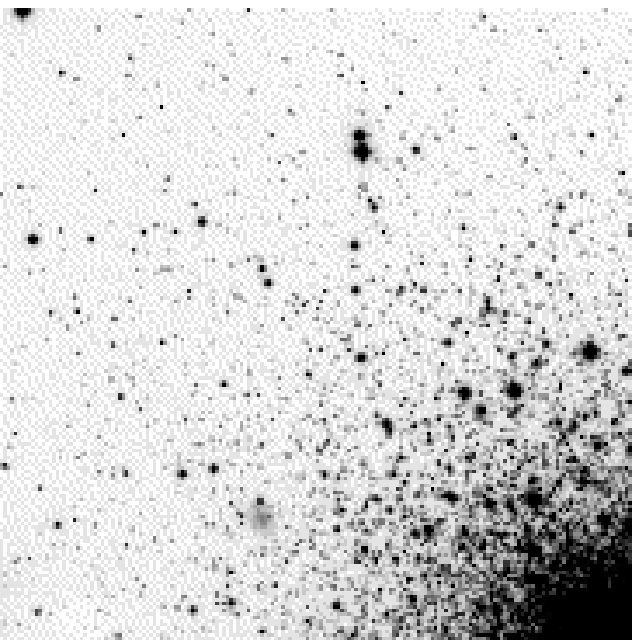,width=6cm}}
\end{tabular}
\caption[]{CMD and covered field for NGC~6341 (M~92)}
\label{ngc6341}
\end{figure*}

\paragraph {\bf NGC~6341 (M~92).} (Fig.~\ref{ngc6341})

M~92 is one of the most metal-poor and one of the best studied globular
clusters in the Galaxy. It was (together with M~3) the first GGC to
be studied down to the TO (Arp, Baum \& Sandage, \cite{arp52},
\cite{arp53}).  Since then, many CMDs have been built for NGC~6341.
The first CCD photometry is presented by Heasley \& Christian
(\cite{heasley86}). They obtain a CMD down to $V=22$ in the $B$ and
$V$ filters, but poorly populated. Another exhaustive work on M~92 is
presented by Stetson \& Harris, who present a deep $B$ and $V$ CMD
with a very well defined MS, but still poorly populated in the evolved
part of the diagram. More recently, Johnson \& Bolte
(\cite{johnson98}) present an excellent $V$ and $I$ diagram of M~92
where the principal sequences are very well defined, but there is
still a small number of stars in the evolved regions. $JHK$ photometry
of this cluster is presented by Davidge \& Courteau
(\cite{davidge99}). Piotto et al. (\cite{piotto97}) present a deep CMD
from HST/WFPC2 extended down to $\sim 0.15M_\odot$. 

Our diagram, with $\sim13900$ stars measured, is well defined and
extends from the RGB tip to about 4 magnitudes below the MS TO.

\begin{figure*}
\begin{tabular}{cc}
\raisebox{-2cm}{\psfig{figure=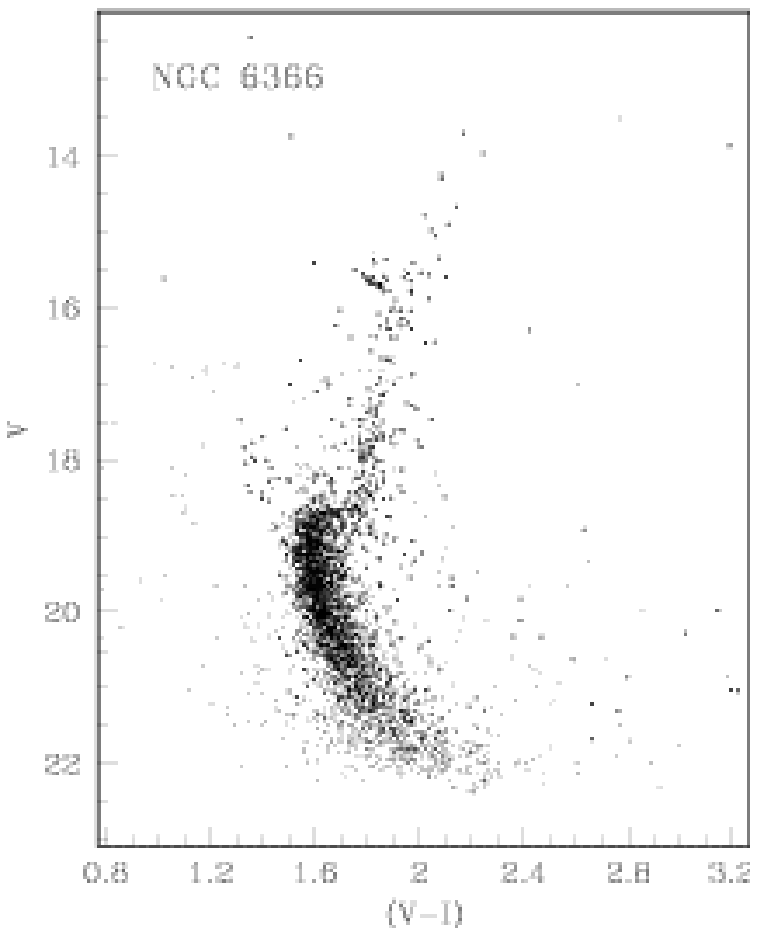,width=8.8cm}} &
\fbox{\psfig{figure=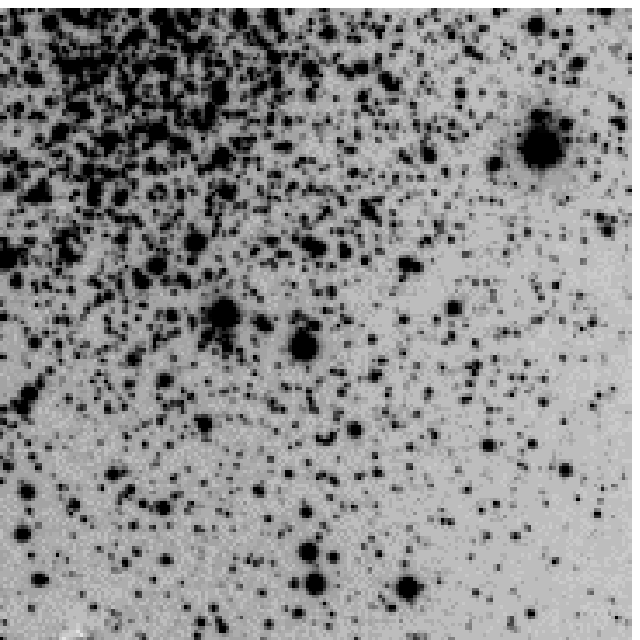,width=6cm}}
\end{tabular}
\caption[]{CMD and covered field for NGC~6366}
\label{ngc6366}
\end{figure*}

\paragraph {\bf NGC~6366.} (Fig.~\ref{ngc6366})

This cluster is somehow peculiar.  While located in the disk, and
being a metal-rich cluster (as M~71 or 47 Tucanae), it has a
kinematics typical of a halo cluster. It is also highly reddened, and
its CMD is affected by some differential reddening.

Alonso et al. (\cite{alonso97}) present the only other CCD $B$ and $V$
CMD existent for this cluster. NGC~6366, together with NGC~5053, are
the only two clusters that were observed under not exceptionally good
seeing conditions.  Still, all the sequences in the CMD can be
identified (apart from the upper RGB), including what seems to be a
well populated blue straggler sequence. The HB is very red, as
expected on the basis of the metallicity, and tilted. We measured a
total of $\sim 5500$ stars for this cluster.

\begin{figure*}
\begin{tabular}{cc}
\raisebox{-2cm}{\psfig{figure=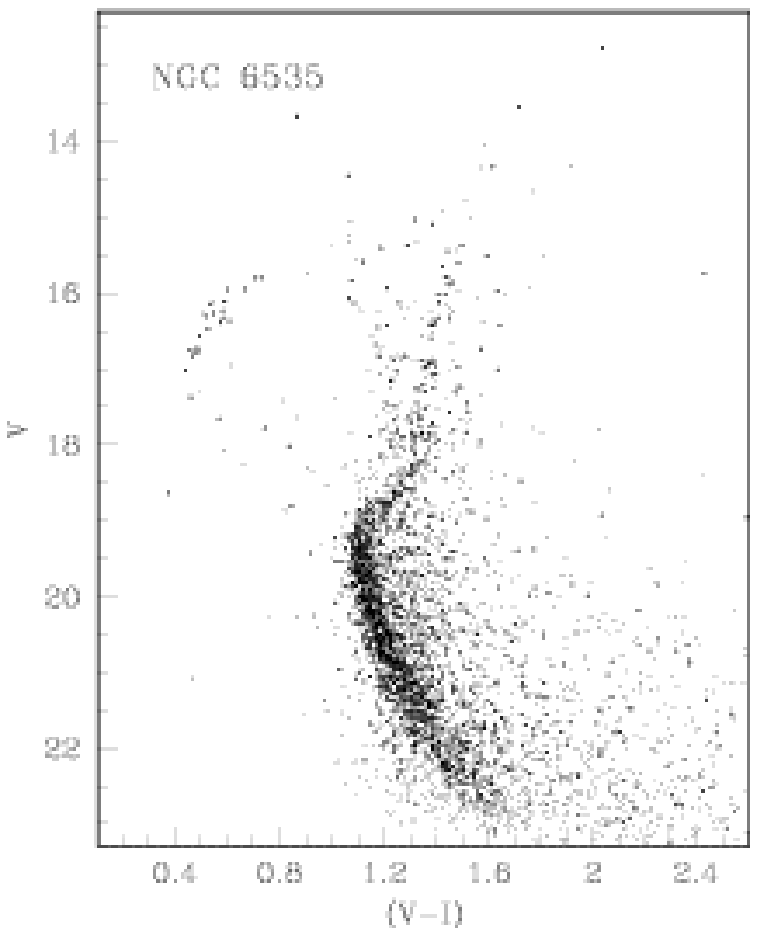,width=8.8cm}} &
\fbox{\psfig{figure=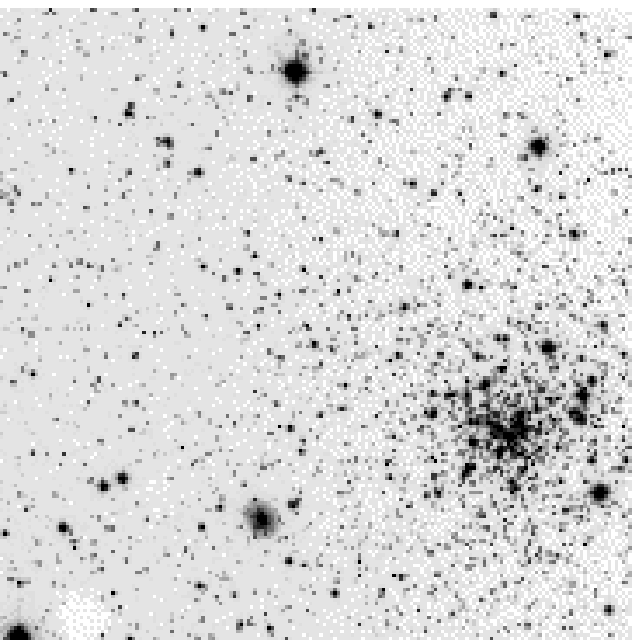,width=6cm}}
\end{tabular}
\caption[]{CMD and covered field for NGC~6535}
\label{ngc6535}
\end{figure*}

\paragraph {\bf NGC~6535.} (Fig.~\ref{ngc6535})

To our knowledge, Sarajedini (\cite{sarajedini94}) has published the
only previous CCD study of this cluster: a $B$ and $V$ CMD down to
$V\sim21$. His stellar population is slightly smaller than ours for
this range of magnitudes (we reach V $\sim23$). NGC~6535 is the least
luminous object of our northern sample, and probably the one with the
smallest number of stars. We measured $\sim7800$ stars for this
cluster. Its RGB is identifiable, but not clearly defined, due also
to field star contamination. Its CMD somehow resembles the CMD of
NGC~6717 (Paper I).

\begin{figure*}
\begin{tabular}{cc}
\raisebox{-2cm}{\psfig{figure=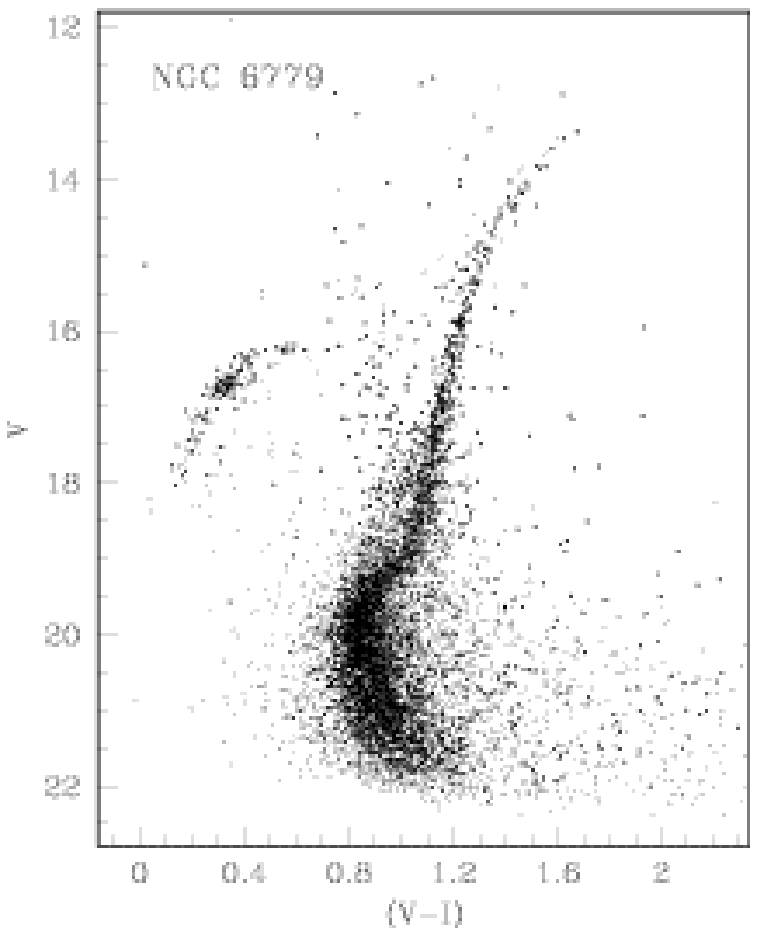,width=8.8cm}} &
\fbox{\psfig{figure=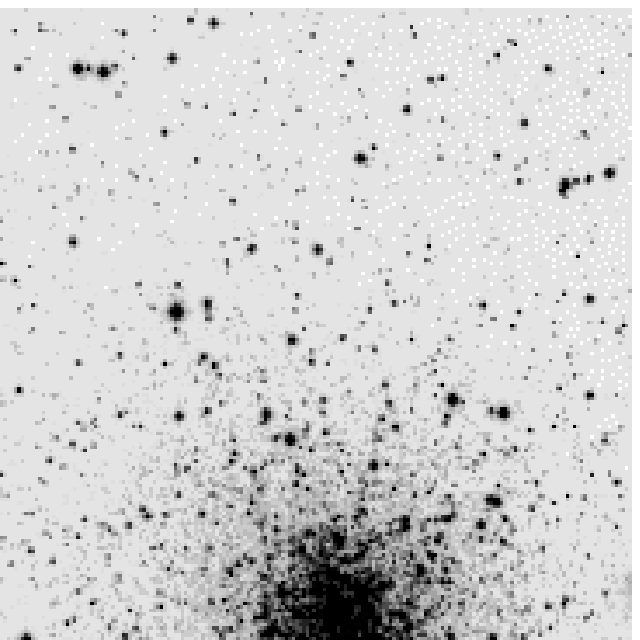,width=6cm}}
\end{tabular}
\caption[]{CMD and covered field for NGC~6779 (M~56)}
\label{ngc6779}
\end{figure*}

\paragraph {\bf NGC~6779 (M~56).} (Fig.~\ref{ngc6779})

We have not found any previous CCD study on this cluster. Our CMD is
well defined, though it is slightly contaminated by
foreground/background stars. The broadening of the SGB-RGB might
suggest the existence of some differential reddening. The distribution
of the stars along the BHB seems to be not homogeneous, with the
possible presence of a gap. The total of measured stars was of
$\sim11300$.

\begin{figure*}
\begin{tabular}{cc}
\raisebox{-2cm}{\psfig{figure=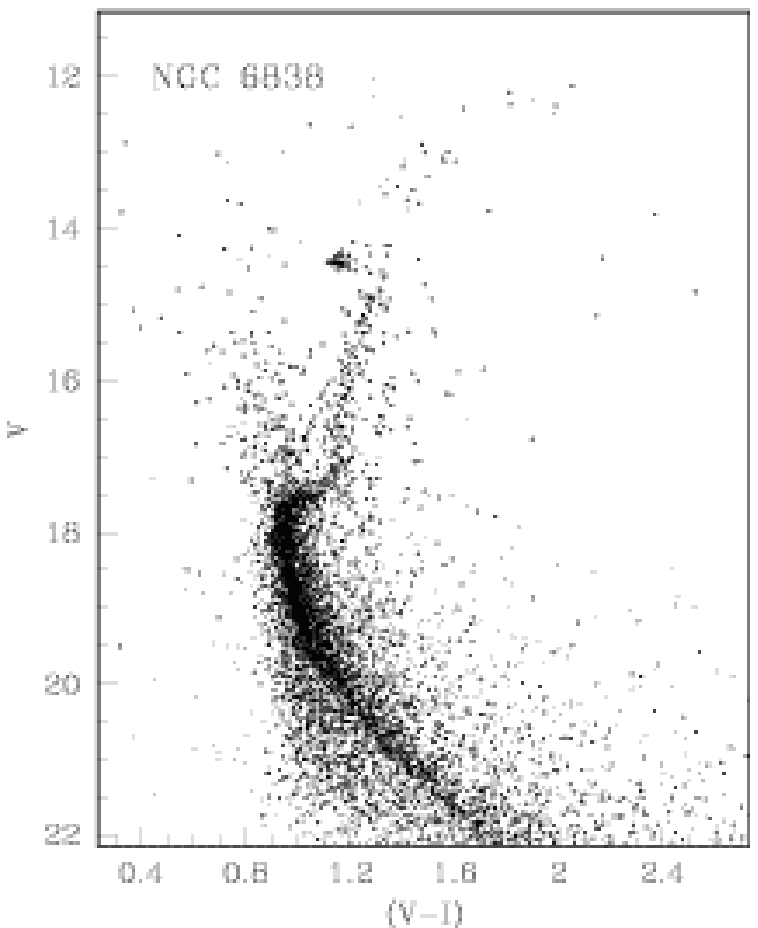,width=8.8cm}} &
\fbox{\psfig{figure=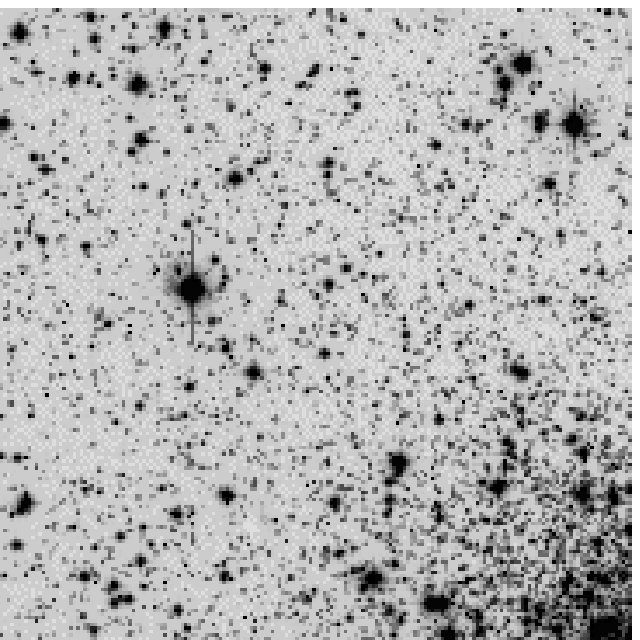,width=6cm}}
\end{tabular}
\caption[]{CMD and covered field for NGC~6838 (M~71)}
\label{ngc6838}
\end{figure*}

\paragraph {\bf NGC~6838 (M~71).} (Fig.~\ref{ngc6838})

As suggested also by its CMD, M71 is a metal rich cluster, similar to
47~Tuc (Paper I). Our CMD is well defined and extends for more than 4
magnitudes below the TO, covering a total of $\sim12500$ stars. The
cluster has only a RHB, and the upper part of the RGB is not very well
defined. This cluster is located close to the Galactic plane, and this
explains the contamination by disk stars clearly visible in the
CMD. It is very bright and relatively nearby.

Despite this, there is no CMD in the literature after Hodder et
al. (\cite{hodder92}). They present a good $B$ and $V$ diagram, less
populated than ours, reaching $V=22$.

Previous CCD studies are in Richer \& Fahlman (\cite{richer88}), who
present $U,B,V$ photometry for the main sequence, down to $V=22$
($U=25$). No evolved stars are present in this work.

\begin{figure*}
\begin{tabular}{cc}
\raisebox{-2cm}{\psfig{figure=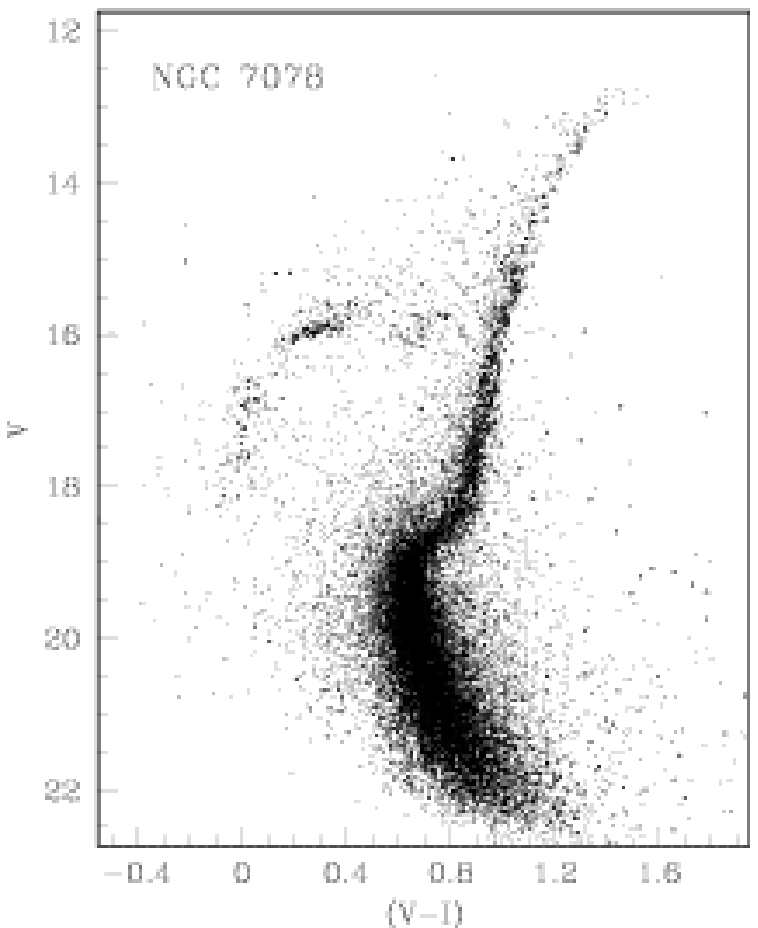,width=8.8cm}} &
\fbox{\psfig{figure=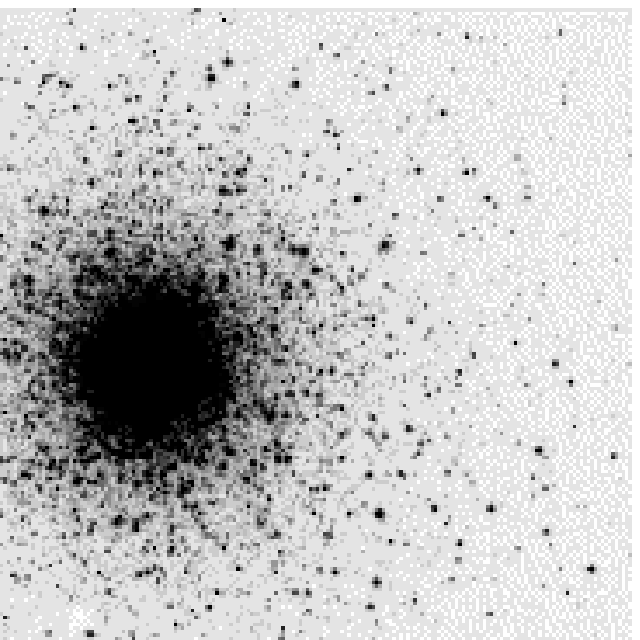,width=6cm}}
\end{tabular}
\caption[]{CMD and covered field for NGC~7078 (M~15)}
\label{ngc7078}
\end{figure*}

\paragraph {\bf NGC~7078 (M~15).} (Fig.~\ref{ngc7078})

This cluster has been extensively studied in the past, both with
groundbased facilities and a large number of HST observations.

HST studies include Stetson \cite{stetson94} and Yanny et al
\cite{yanny94}, were a CMD of the cluster center is presented. The CMD
does not arrive to the MS TO, and is quite disperse. Conversely, Sosin
\& King \cite{sosin97} and Piotto et al. \cite{piotto97}) present and
extraordinarily well defined MS, but no evolved stars are present.

The most recent ground-based study is the composite CMD of Durrell \&
Harris (\cite{durrell93}) based on CCD data from two telescopes. This
is the kind of problem that we try to avoid with the present catalog.

Our diagram is well populated ($\sim27000$ stars) from the RGB tip
down to $V=22.5$. The CMD features are better identifiable when a
radial selection, avoiding the clusters center, is done. The CMD in
Fig.~\ref{ngc7078} gives the visual impression that there are three
distinct groups of stars in the HB. The third possible group, on the
red side of the RR Lyrae gap is surely a statistical fluctuation in
the distribution of the RR Lyrae magnitudes and colors at random
phase. It is present neither in the CMDs of M15 in the above quoted
works nor in our CMD of a larger stellar sample, with more accurate
photometry from our HST data base.

\begin{acknowledgements}
This paper has been partially supported by the Ministero della Ricerca
Scientifica e Tecnologica under the program ``Treatment of large format
astronomical images'' and by CNAA.
\end{acknowledgements}


\end{document}